\documentclass[5p, twocolumn]{elsarticle}
\usepackage{amsmath, amsfonts, amssymb}
\biboptions{sort&compress}
\journal{Journal of Magnetism and Magnetic Materials}
\begin{document}
\begin{frontmatter}

\title{Ising model on a square lattice with second-neighbor and third-neighbor interactions} 
\author[label1]{F.A.~Kassan-Ogly} 
\ead{felix.kassan-ogly@imp.uran.ru}
\author[label2]{A.K.~Murtazaev}
\author[label1]{A.K.~Zhuravlev} 
\author[label2]{M.K.~Ramazanov} 
\author[label1]{A.I.~Proshkin} 
\address[label1]{Institute of Metal Physics, Ural Division,
Russian Academy of Sciences, ul. S.Kovalevskoi 18, Ekaterinburg, 620990, Russia}
\address[label2]{Institute of Physics, DSC,
Russian Academy of Sciences, ul. Yaragskogo 94, Makhachkala, Daghestan, 367003, Russia}

\begin{abstract}
We studied the phase transitions and magnetic properties of the Ising model on a square lattice by the replica Monte Carlo method and by the method of transfer-matrix, the maximum eigenvalue of which was found by Lanczos method. The competing exchange interactions between nearest neighbors $J_{1}$, second $J_{2}$, third neighbors $J_{3}$ and an external magnetic field were taken into account. We found the frustration points and expressions for the frustration fields, at crossing of which cardinal changes of magnetic structures (translational invariance changes discontinuously) take place. A comparative analysis with 1D Ising model was performed and it was shown that the behavior of magnetic properties of the 1D model and the 2D model with $J_{1}$ and $J_{3}$ interactions reveals detailed similarity only distinguishing in scales of magnetic field and temperature.
\end{abstract}

\end{frontmatter}

\section{Introduction}
It has been over 70 years since the publication of Onsager celebrated solution \cite{ons} of the Ising model on a square lattice with nearest-neighbor (nn) interactions. This solution has served as a cornerstone for modern theories and as a testing ground for many approximate theoretical approaches. Since that time, there have been very few other systems for which exact solutions have been found. Among them the exact solutions of the Ising model on other 2D lattices: a triangular \cite{wan}, a honeycomb \cite{hou}, and a kagome lattice \cite{kan} have been found, but again only with nearest-neighbor (nn) interactions. The slight alterations of the original model such as the addition of next-nearest-neighbor (nnn) interactions or an external magnetic field are no longer exactly soluble by presently available theoretical approaches. Naturally without an exact solution different approximate methods such as mean-field approximations, series expansions \cite{dal, rap, oit, lee}, calculating the interface free energy \cite{mul}, real-space renormalization-group techniques \cite{nie, sub, sch, kin}, finite-size scaling of the transfer matrix \cite{que, nig}, the Fisher zeros of the partition function \cite{mon, kim}, and various kinds of Monte Carlo simulations \cite{das, lan, bin, lanbin, moe, lanwan, mal, kal1, kal2, jin} were developed up to recent years. 

The overwhelming majority of papers on a square lattice in the Ising model are devoted to multifarious topical problems such as determining and elucidating the phase diagrams of magnetic structures, the order and the universality class of the phase transitions; finding the multicritical points and critical lines that separate the different phases; determining the frustration lines and points that result in the appearance of Quantum Phase Transitions; calculating the critical exponents, etc. Nevertheless, certain issues have remained controversial either due to different theoretical approaches or, more often, due to insufficient sizes of a lattice subjected to numerical calculations.

However, despite the significant amount of effort made in many years, a number of related problems still remain essentially untouched, in particular, the behavior of magnetization, as a function of an external magnetic field and (or) the temperature.

The aim of the present paper is just the study of magnetic properties of the Ising model on a square lattice in comparison with those on a one-dimensional (linear chain) lattice.

\begin{figure}
	\begin{center}
	\includegraphics[keepaspectratio,width= 7 cm]{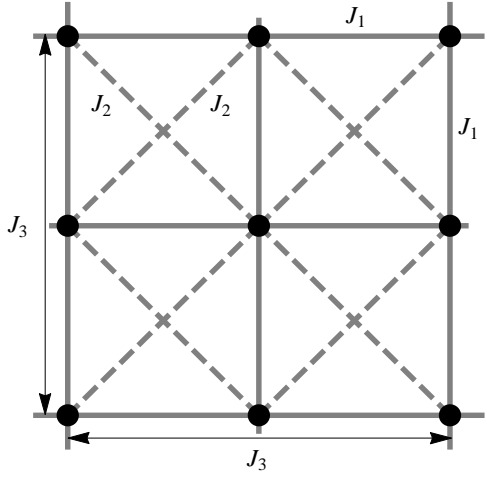}
	\caption{The nearest-, second-, and third-neighbor interactions on a 		square lattice in the Ising model.}
	\end{center}
	\label{fig:lattice}
	\end{figure}

The Ising model on a $N = L \times L$ square lattice (or on a linear chain with $L$ sites) is described by the Hamiltonian:
\begin{equation}
\label{eq:ham}
{\cal H} =J_1 \sum_{nn} S_i S_j + J_2 \sum_{nnn} S_i S_k+ J_3 \sum_{nnnn} S_i S_l-H\sum_{i} S_i,
\end{equation}
where all the exchange interactions $J_1$, $J_2$ and $J_3$ are positive (antiferromagnetic); nn, nnn, and nnnn stand, respectively, for all next-neighbor pairs, second-neighbor pairs, and third-neighbor pairs; $S_i = \pm 1$; $H$ is an external magnetic field. Here, all fields, exchange interactions, and temperatures are given in units of $J_1$, unless otherwise stated. We have kept $J_1 =1$ in all the calculations reported in this work. The exchange interactions on a square lattice are shown in Fig.~\ref{fig:lattice}. 

On a square lattice, we will numerically calculate the magnetization, entropy, and heat capacity as functions of the temperature and an external magnetic field according to the conventional formulas:
\begin{align}
\label{eq:main}
F&=-\frac{T}{N} \ln Z & M&= - \frac {\partial {F}}{\partial H} \\
S&= - \frac {\partial {F}}{\partial T} & C&= - T \frac {\partial^2{F}}{\partial T ^2},
\end{align}
where $Z$ is the partition function, $F$ is the free energy, and $N$ is the number of sites on a square lattice.
On a one-dimensional (linear chain) lattice, we will perform calculations using the exact analytical solution for the maximum eigenvalue $\lambda_{\text{max}}$ of the Kramers-Wannier transfer-matrix, obtained in \cite{kas}. In this case, the magnetization and entropy are expressed solely in terms of $\lambda_{\text{max}}$ as follows:
\begin{align}
\label{eq:eig}
M&=\frac{T}{\lambda_{\text{max}}} \frac {\partial {\lambda_{\text{max}}}}{\partial H},\\
S&=\ln \lambda_{\text{max}} + \frac{T}{\lambda_{\text{max}}} \frac {\partial {\lambda_{\text{max}}}}{\partial T},\\
C&=2\frac{T}{\lambda_{\text{max}}} \frac {\partial {\lambda_{\text{max}}}}{\partial T}+T^2 \frac{\partial}{\partial T}\frac {1}{\lambda_{\text{max}}} \frac {\partial {\lambda_{\text{max}}}}{\partial T}. 
\end{align}

The maximum transfer-matrix eigenvalue $\lambda_{\text{max}}$ is given by 
\begin{multline}
\label{eq:max}
\lambda_{\text{max}} = \frac{\sqrt {a^2-4b+4y}-a}{4}+\\
\sqrt {\left(\frac{{a^2-4b+4y}}{4}\right)^2-\frac{y}{2}-\frac{2c-ya}{2\sqrt {a^2-4b+4y}}},
\end{multline}
where
\begin{equation}
\label{eq:coef}
\begin{split}
y &= \sqrt[3]{\sqrt{Q}-\frac{q}{2}}+\sqrt[3]{-\sqrt{Q}-\frac{q}{2}} +\frac{b}{3},\enspace Q= \frac{p^3}{27}+\frac{q^2}{4},\\
p&=-\frac{b^2}{3}+ac-4d, \enspace q=-\frac{2b^3}{27}+\frac{bac}{3}-\frac{8bd}{3}-a^2d-c^2,\\
a&=-2 \exp \left( \frac{-J_1-J_2}{T} \right) \cosh{\frac{H}{T}},\\
b&=-2 \exp \left( \frac{-2J_2}{T} \right) \sinh{\frac{2J_2}{T}},\\
c&=-4 \exp \left( \frac{J_1-J_2}{T} \right) \sinh{\frac{2J_2}{T}} \cosh{\frac{H}{T}},\\
d&=-4\sinh^2{\frac{2J_2}{T}}.
\end{split}
\end{equation}

\section{Nearest-neighbor interaction}

\begin{figure}
\begin{center}
\includegraphics[keepaspectratio,width= 7 cm]{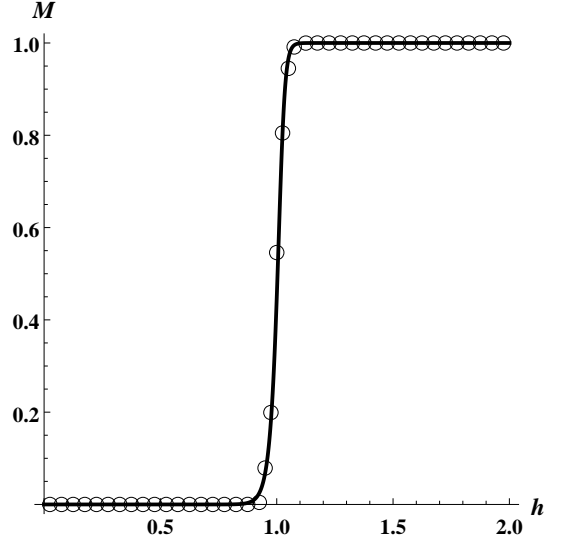}
\caption{Magnetizations of 1D chain and 2D square lattice. Ising model with only nearest-neighbor interactions.}
\end{center}
\label{fig:2}
\end{figure}
	
\begin{figure*}
\includegraphics[keepaspectratio,width= 18 cm]{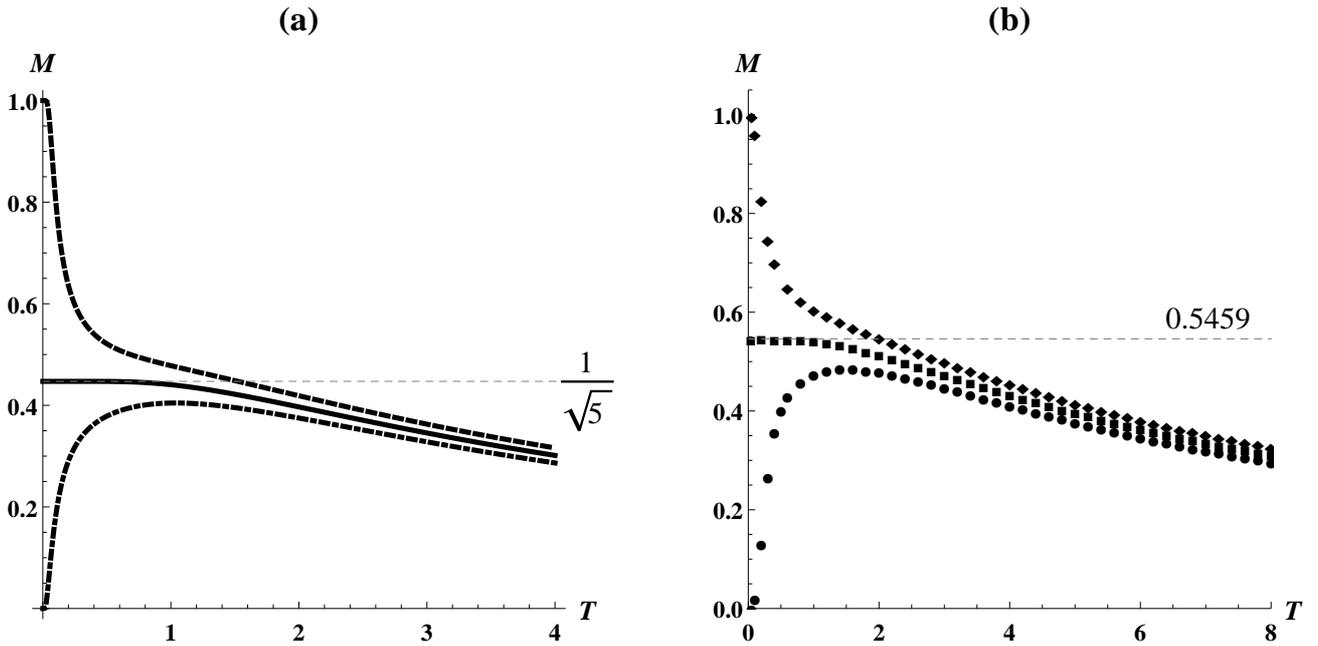}
\caption{Temperature dependence of magnetizations of 1D chain (a) and 2D square lattice (b) at: $H = H_{\text{fr}}$ (middle curves), $H = 0.95H_{\text{fr}}$ (lower curves), and $H = 1.05H_{\text{fr}}$ (upper curves).}
\label{fig:3}
\end{figure*}
	
Let us first consider the simplest case, when $J_2$ and $J_3$ are equal to zero in the Hamiltonian~(\ref{eq:ham}). Figure~\ref{fig:2} shows the magnetizations as functions of rescaled magnetic field $h=H/z$, where $z$ is the number of nearest neighbors on 1D chain ($z=2$) or on 2D square lattice ($z=4$). The magnetizations are calculated at $T=0.05$ on 1D chain (solid line) and at $T=0.1$ on a square lattice (open circles). It can be seen that there exist frustration fields on both a linear chain and square lattice that are, respectively, equal to $H_{\text{fr.1D}}=2J$ and $H_{\text{fr.2D}}=4J$, and that can be expressed uniformly as $H_{\text{fr}}=zJ$. It should be noted that on such a scale the magnetizations are almost indistinguishable from one another.
        
In Fig.~\ref{fig:3}, we show plots of the magnetization on a linear chain and square lattice as a function of temperature at different values of magnetic field: at frustration fields, above, and below them. It is again seen that the magnetization behavior is very similar. At $T\rightarrow 0$ and $H>H_{\text{fr}}$ the magnetization approaches unity, at $H<H_{\text{fr}}$ it vanishes, and at $H=H_{\text{fr}}$ it tends to constant values, although different.

\begin{figure}
	\includegraphics[keepaspectratio,width= 8.5 cm]{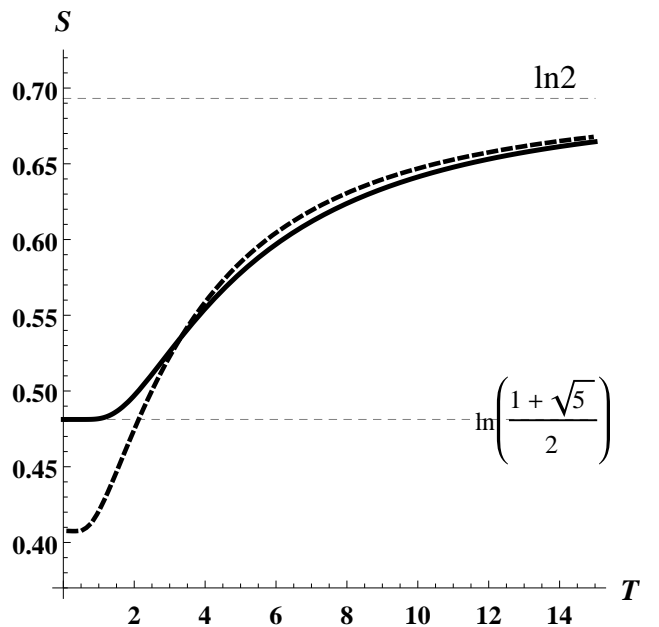}
	\caption{Entropy of Ising model with only nearest-neighbor interactions as a function of temperature at frustration fields on 1D chain (solid line) and 2D square lattice (dashed line).}
	\label{fig:4}
	\end{figure}

Our calculations of the entropy have shown that at $T\rightarrow 0$ and $H=H_{\text{fr}}$ on both the linear chain and square lattice it tends to non-zero value, which is shown in Fig.~\ref{fig:4}. The non-zero entropy value at $T=0$ indicates vanishing of the transition temperature at these fields (see, for example, \cite{yin}) and gives evidence that these fields are really the frustration points. The calculations of heat capacity at and nearby the frustration fields also confirm this assertion. The suppression of the transition temperature on a square lattice at $H_{\text{fr.1D}}=4 J$ was also corroborated by M\"{u}ller-Hartmann and Zittartz in \cite{mul}, where they put forward a considerably simple conjecture for the relation between the transition temperature and an external magnetic field
\begin{equation}
\label{eq:mul}
\cosh{\left( \dfrac{H}{T_c}\right) } = \sinh^2{\left( \dfrac{2J}{T_c} \right) }.
\end{equation}

The relation~(\ref{eq:mul}) had been more or less successfully confirmed by subsequent Monte-Carlo simulations \cite{kin, bin} and real space renormalization group method \cite{sch}.
	
\section{Nearest-neighbor and second-neighbor interactions}
Having seen in the previous section many similarities in the behavior of magnetic properties between 1D chain and 2D square lattice let us consider the second case, when $J_2\neq 0$ and $J_3=0$ in the Hamiltonian with a hope to find new similar features. The majority of papers are particularly devoted to this case. However, despite of a plethora of articles, the calculations of the magnetization dependence on temperature and magnetic field are virtually lacking. 
	
	\begin{figure}
	\includegraphics[keepaspectratio,width= 8.5 cm]{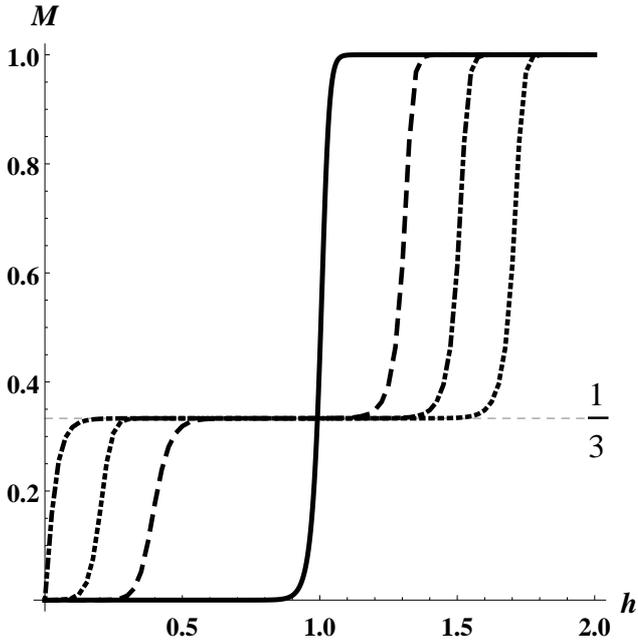}
	\caption{Magnetizations of Ising model with nearest-neighbor and second-neighbor interactions on a linear chain at: $R = 0$ (solid curve), $R = 0.3$ (dashed curve), $R = 0.5$ (dot-and-dash curve), $R = 0.7$ (dotted curve). $T=0.05$.}
	\label{fig:5}
	\end{figure}
	\begin{figure}
	\includegraphics[keepaspectratio,width= 8.5 cm]{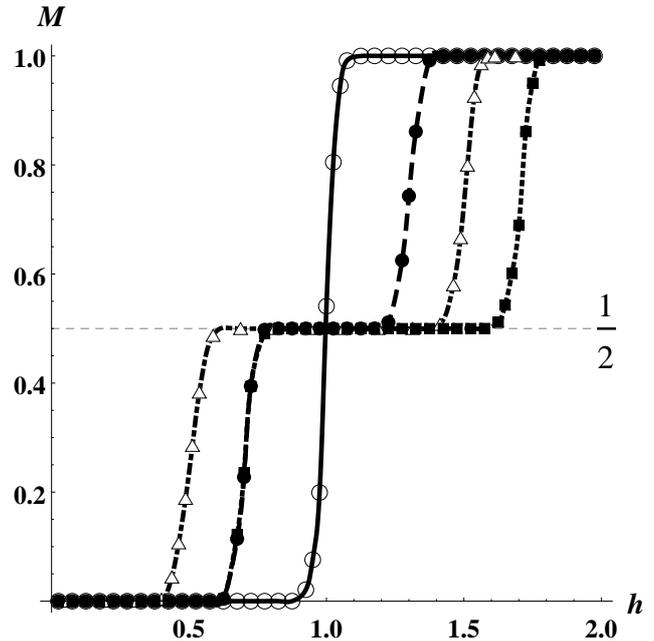}
	\caption{Magnetization of Ising model with nearest-neighbor and second-neighbor interactions on a square lattice at: $R = 0$ (solid curve), $R = 0.3$ (dashed curve), $R = 0.5$ (dot-and-dash curve), $R = 0.7$ (dotted curve). $T=0.1$.}
	\label{fig:6}
	\end{figure}
	
\begin{figure}
	\includegraphics[keepaspectratio,width= 8.5 cm]{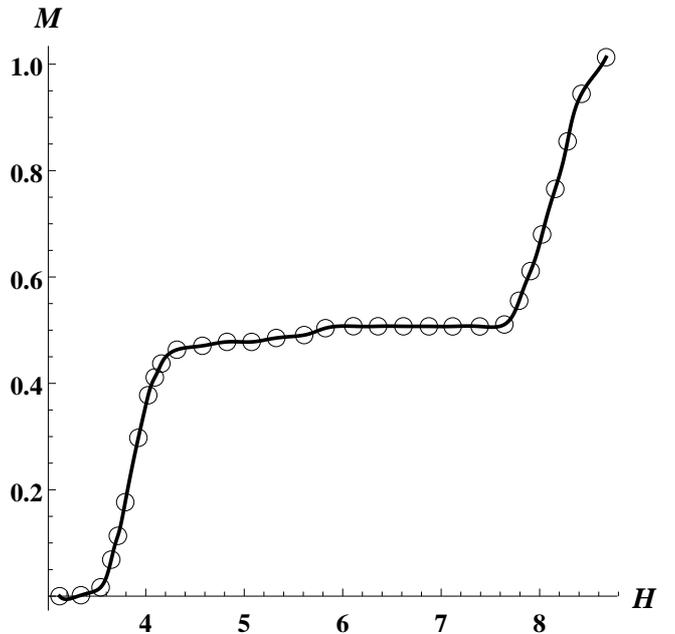}
	\caption{Magnetization of Ising model with nearest-neighbor and second-neighbor interactions on a square lattice at $R = 1.0$, \cite{bin}.}
	\label{fig:7}
	\end{figure}

In Figs.~\ref{fig:5}~and~\ref{fig:6}, we show plots of the magnetization on a linear chain and a square lattice as a function of rescaled magnetic field $h=H/z$ at different values of the ratio between second-neighbor $J_2$ and nearest-neighbor $J_1$  interactions ($R=J_2/J_1 = 0,\,0.3,\,0.5$ and $0.7$). We have found a calculated magnetization curve for $R=1$ and at $T=0.273$ in \cite{bin} (see Fig.~\ref{fig:7}).

Analyzing Figs.~\ref{fig:5}~and~\ref{fig:6}, we can notice some similarities in the magnetization behavior on both the 1D chain and 2D square lattice, namely, the appearance of two frustration fields (lower and upper), a plateau at intermediate values of a magnetic field, and an antiferromagnetic type of magnetization increasing at very low fields (magnetic susceptibility vanishes at $H\rightarrow 0$ for almost all the curves). 

However along with this, several crucial dissimilarities are revealed in the magnetizations behavior. First of all, the heights of plateaus are different; $1/2$ of saturation magnetization on the square lattice, and $1/3$ on the 1D chain. Second, at the frustration point of the interactions ratio $R=1/2$ at low magnetic fields the magnetization has a ferromagnetic type (at $H\rightarrow 0$ the susceptibility goes to infinity). Third, the frustration fields for a linear chain and a square lattice do not coincide even in the rescaled form.
\begin{equation}
\label{eq:fr1}
\begin{split}
H_{\text{fr.1D}}^{\text{low}}&=\begin{cases}
z J_1-2 z J_2,&\text{at $0<R<0.5$;}\\
z J_2-\dfrac{z}{2}J_1,&\text{at $R>0.5$.} 
\end{cases}\\
H_{\text{fr.1D}}^{\text{upp}}&=z J_1+ z J_2.
\end{split}
\end{equation}
\begin{equation}
\label{eq:fr2}
\begin{split}
H_{\text{fr.2D}}^{\text{low}}&=\begin{cases}
z J_1-z J_2,&\text{at $0<R<0.5$;}\\
z J_2,&\text{at $R>0.5$.} 
\end{cases}\\
H_{\text{fr.2D}}^{\text{upp}}&=z J_1+ z J_2.
\end{split}
\end{equation}
   
    \begin{figure}
	\includegraphics[keepaspectratio,width= 8.5 cm]{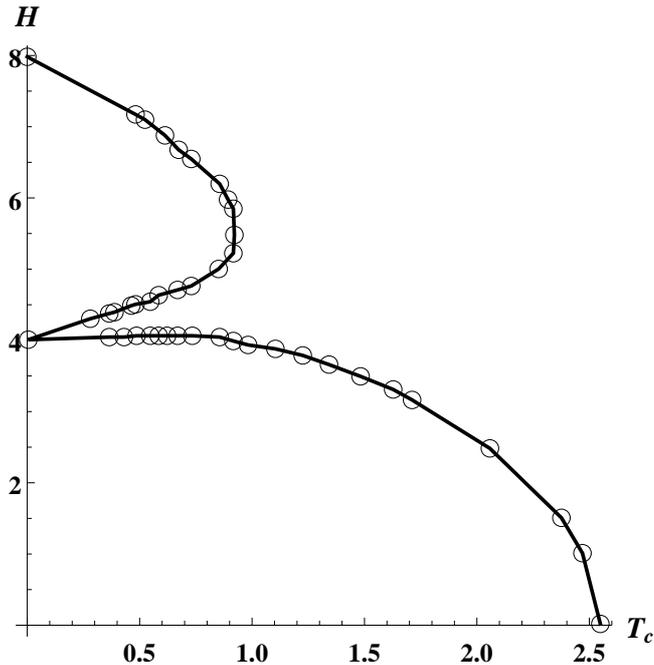}
	\caption{Phase diagram for Ising square lattice with antiferromagnetic nearest-neighbor and second-neighbor interactions in a magnetic field for $R=1.0$, \cite{yin}.}
	\label{fig:8}
	\end{figure} 
	\begin{figure}
	\includegraphics[keepaspectratio,width= 8.5 cm]{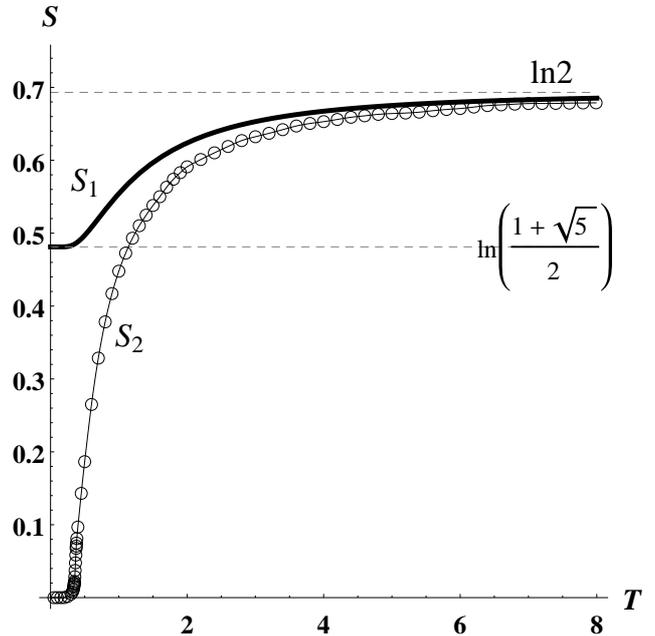}
	\caption{Temperature dependences of entropies in Ising model with nearest- and second-neighbor interactions on 1D chain (solid line) and square lattice (open circles) at the frustration point $R = 0.5.$}
	\label{fig:9}
	\end{figure}
	\begin{figure}
	\includegraphics[keepaspectratio,width= 8.5 cm]{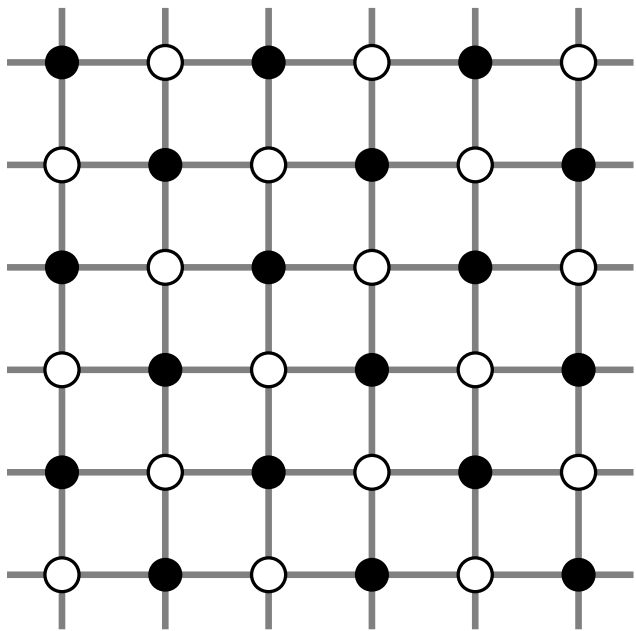}
	\caption{The N\'{e}el structure.}
	\label{fig:10}
	\end{figure} 

In the Ising model with antiferromagnetic nearest- and second-neighbor interactions on a square lattice, the frustration fields~(\ref{eq:fr2}) were for the first time determined by Binder and Landau in \cite{bin}, in which they attempted to calculate the field dependence of the transition temperature for various values of the ratio $R$, using the mean-field approximation. In the subsequent paper \cite{yin} after elaborated Monte Carlo simulations, the authors obtained the correct field dependence of the transition temperature. The phase diagram (or the transition curve) for the Ising square lattice with antiferromagnetic nearest- and second-neighbor interactions in a magnetic field for $R=1$ is shown in Fig.~\ref{fig:8}.

	\begin{figure*}
	\includegraphics[keepaspectratio,width= 18 cm]{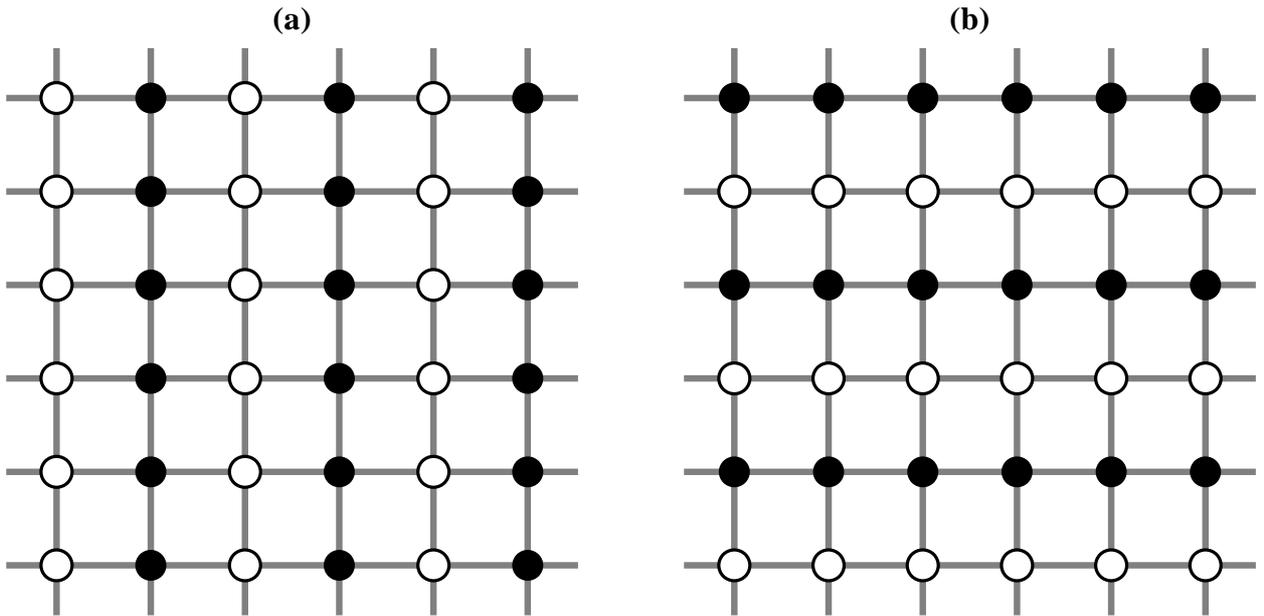}
	\caption{Superantiferromagnetic structure.}
	\label{fig:11}
	\end{figure*}
	
The calculations of the entropy as a function of temperature at the frustration point of the interactions ratio $R=1/2$ on a linear chain and a square lattice plotted in Fig.~\ref{fig:9} also shows quite different behavior. At $T\rightarrow 0$, the entropy on a linear chain tends to a constant value $\ln \frac{1+\sqrt 5}{2}$  that is the logarithm of golden ratio. At the same time, on a square lattice at $J_1\neq 0$ and $J_3=0$ the entropy vanishes, nevertheless the transition temperature is suppressed to zero, which follows from the heat capacity calculations.

Let us now discuss the ground state magnetic structures that can be found from energy arguments, as was first done by Fan and Wu \cite{fan}. When the nearest-neighbor interaction is strong enough $J_1>2J_2$ ($R<1/2$) it determines the magnetic ordering. In this case, the energy per site is equal to $E_{\text{N\'{e}el}}=-2J_1+2J_2$, and the magnetic structure is known as antiferromagnetic or the N\'{e}el-ordered state depicted in Fig.~\ref{fig:10}. Hereafter solid and open circles represent the spin states $S_i=\pm 1$.

In the opposite case, when the second-neighbor interaction is strong $J_2>J_1/2$ ($R>1/2$) the energy per site only depends on $J_2$ and equals $E_{\text{SAF}}=-2J_2$. The magnetic structure, shown in Fig.~\ref{fig:11}, was names by Fan and Wu \cite{fan} "superantiferromagnetic". In many other subsequent papers it is referred to as SAF, "striped" or "layered" structure. It looks like vertical chains are ordered in antiferromagnetic manner (Fig.~\ref{fig:11}a) or like horizontal chains are ordered antiferromagnetically (Fig.~\ref{fig:11}b) with keeping translational invariance (as if some ferromagnetic interaction is present) in the direction along one side of a square and with doubled period in the other. But this impression is erroneous, and the structure should be perceived as antiferromagnetic ordering (determined by strong $J_2$) along the both diagonals.
	
	\begin{figure}
	\includegraphics[keepaspectratio, width= 8.5 cm]{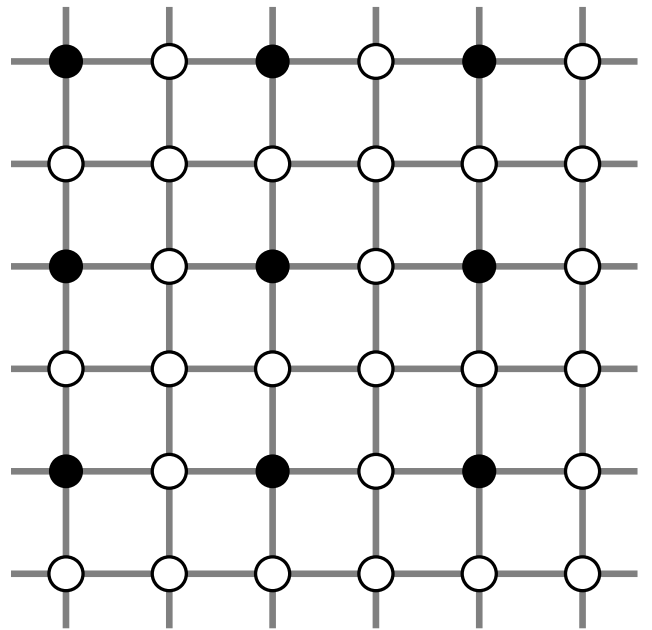}
	\caption{Magnetic structure in intermediate field (between the lower and upper frustration fields) in a model with nearest-neighbor and second-neighbor interactions.}
	\label{fig:12}
	\end{figure}
	  
The N\'{e}el and superantiferromagnetic phases are separated by the frustration point at $J_2=J_1/2$ ($R=1/2$), at which the transition temperature is suppressed to zero. Since at $J_2=J_1/2$ energies of the N\'{e}el ($E_{\text{N\'{e}el}}$) and SAF ($E_{\text{SAF}}$) structures are equal, a single translation of any horizontal chain in Fig.~\ref{fig:11}a or any vertical chain in Fig.~\ref{fig:11}b costs no energy. Thus, at the frustration point the ground state is composed of $4\cdot 2^L$ configurations and hence degenerate on the order of $2^L$ (but not on the order of $2^N$) so that the entropy vanishes at $T=0$.

The ground state structure in an external magnetic field below the lower frustration field (\ref{eq:fr2}) coincides either with the N\'{e}el (at $R<1/2$), or with SAF structure (at $R>1/2$). Above the upper frustration field the structure, naturally, is ferromagnetic. In between the lower and upper frustration field, the structure acquires the form shown in Fig.~\ref{fig:12}, in which open circles correspond to spins aligned along an external magnetic field. The obtained structure has a four-fold symmetry and is composed of alternate ferromagnetic and simple antiferromagnetic chains along both sides of a square. 
	
	\begin{figure}
	\includegraphics[keepaspectratio,width= 8.5 cm]{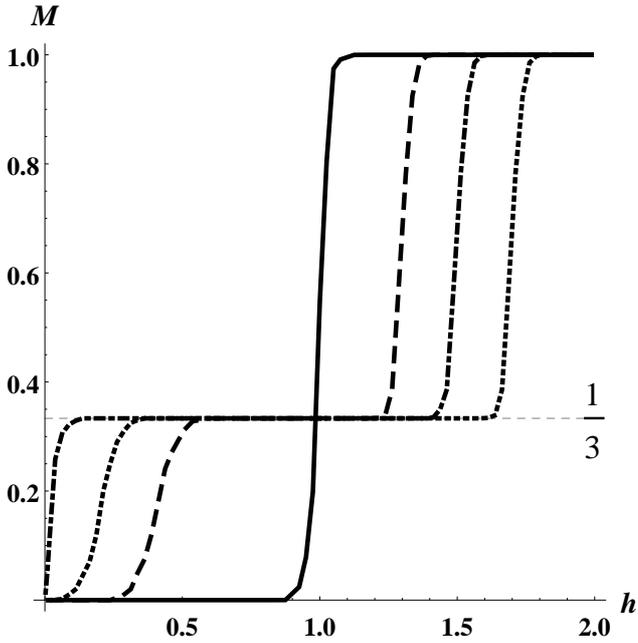}
	\caption{Magnetizations of Ising model with nearest-neighbor and third-neighbor interactions on a square lattice at: $R' = 0$ (solid curve), $R' = 0.3$ (dashed curve), $R' = 0.5$ (dot-and-dash curve), $R'=0.7$ (dotted curve), $T=0.1$.}
	\label{fig:13}
	\end{figure}
	
\section{Nearest-neighbor and third-neighbor interactions}
	\begin{figure}
	\includegraphics[keepaspectratio,width= 8.5 cm]{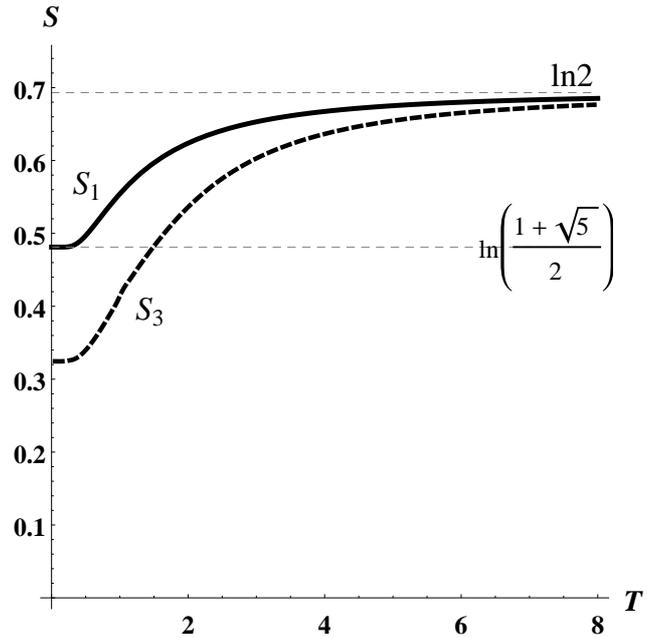}
	\caption{Temperature dependence of entropy in Ising model with nearest-neighbor and second-neighbor interactions on 1D chain (solid line) and with nearest-neighbor and third-neighbor interactions on a square lattice (dashed line) at frustration points $R=0.5$.}
	\label{fig:14}
	\end{figure} 
Let us now consider the magnetic properties of the Ising model on a square lattice in the third case, when $J_3\neq 0$ and $J_2=0$ in the Hamiltonian (\ref{eq:ham}). In this case, our calculations of the magnetization on a square lattice as a function of rescaled magnetic field $h=H/z$ at four values of the ratio between third-neighbor $J_3$ and nearest-neighbor $J_1$ interactions ($R'= J_3/J_1=0,\,0.3,\,0.5$ and $0.7$) are plotted in Fig.~\ref{fig:13}. All of them are calculated at $T=0.1$. A comparison of Figs.~\ref{fig:13}~and~\ref{fig:5} shows a striking similarity between the magnetizations on a linear chain and a square lattice. Keeping in mind different scales of temperature and magnetic field the matched curves are almost indistinguishable. It should be emphasized that the obtained frustration fields on a square lattice, when $J_3\neq 0$ and $J_2=0$, exactly duplicate those on a linear chain (\ref{eq:fr1}). 

Figure~\ref{fig:14} shows the calculation results of entropy dependence on temperature at the frustration points $R=R'=1/2$ on a linear chain and a square lattice in the case, when $J_3\neq 0$ and $J_2=0$. The entropies on both a linear chain and a square lattice display similar tending to nonzero values at $T\rightarrow 0$ in contrast to dissimilar behavior in the previous case, when $J_2\neq 0$ and $J_3=0$ (compare with Fig.~\ref{fig:9}). 

	\begin{figure*}
	\includegraphics[keepaspectratio,width= 18 cm]{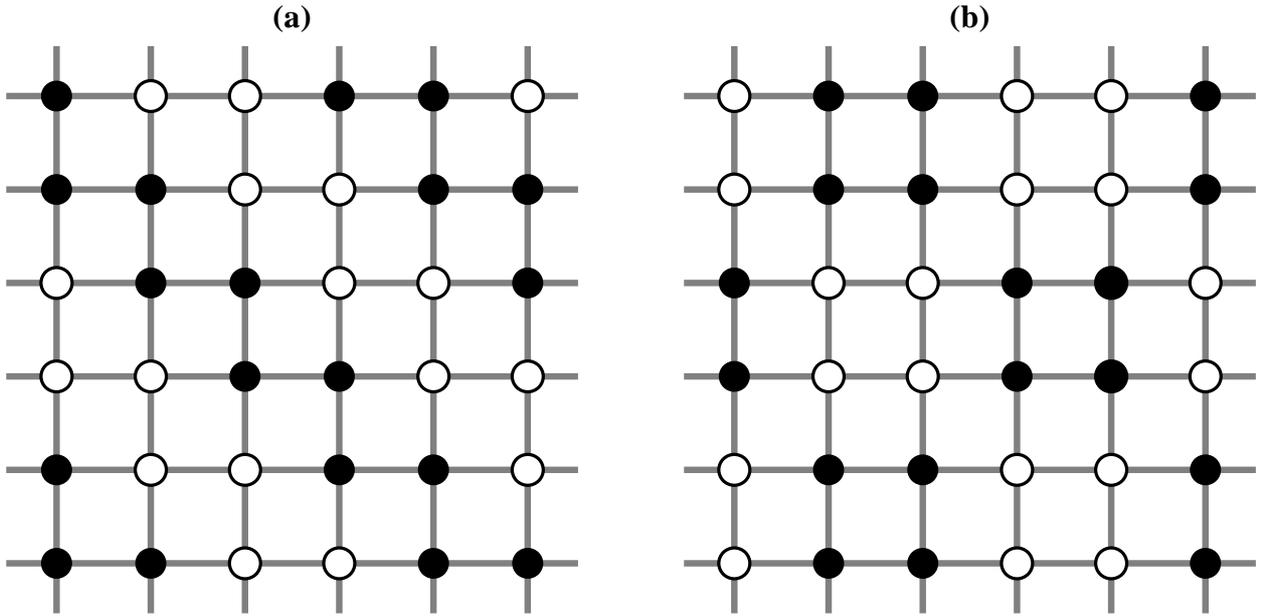}
	\caption{Quadruple structures in Ising model with nearest-neighbor and third-neighbor interactions on a square lattice at $R'>0.5$.}
	\label{fig:15}
	\end{figure*} 

In the case, when $J_3\neq 0$ and $J_2=0$ and at strong $J_1>2J_3$ ($R'<1/2$) the magnetic structure with energy equal to $E=-2J_1+2J_3$ again is the N\'{e}el-ordered one (Fig.~\ref{fig:10}). When the third-neighbor interaction is strong $J_3>1/2J_1$ ($R'>1/2$) the energy per site only depends on $J_3$ and is equal to $E=-2J_3$. Figure~\ref{fig:15} shows two obtained structures of different symmetry. Notwithstanding this difference, they both have the same energy and are composed in a similar way, namely, as alternate sequence of vertical and (or) horizontal chains of $++--++--$ type. We here have the translational invariance with quadruple period along both sides of a square. These structures, the N\'{e}el and quadruple one, are separated by the frustration point at $J_3=1/2J_1$ ($R'=1/2$), at which the transition temperature is suppressed to zero and the entropy, as the calculations showed, tends to a non-zero value at $T\rightarrow 0$ (see Fig.~\ref{fig:14}). The suppression of transition temperature also is corroborated by the heat capacity calculations at and nearby the frustration point $R'=1/2$.

\begin{figure}
	\includegraphics[keepaspectratio,width= 8.5 cm]{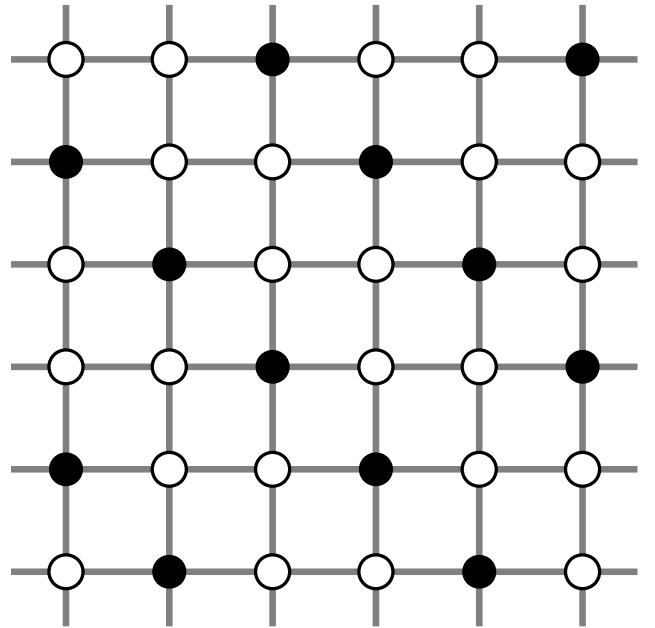}
	\caption{Magnetic structure with triple translational periods at intermediate field (between the lower and upper frustration fields) in a model with nearest-neighbor and third-neighbor interactions.}
	\label{fig:16}
	\end{figure}

In an external magnetic field below the lower frustration field (\ref{eq:fr1}), the ground state structure coincides either with the N\'{e}el (at $R'<1/2$), or with quadruple structure (at $R'>1/2$). Above the upper frustration field the structure, naturally, is ferromagnetic. In between the lower and upper frustration field, the structure acquires the form shown in Fig.~\ref{fig:16}. It should be noted that the obtained structure is composed of alternate vertical and horizontal chains with tripled translational period of $++-++-$ type along both sides of a square, similar to the 1D lattice.
	
\section{Conclusions}
In this paper we have considered the Ising model with competing nearest-neighbor, second-neighbor, and third-neighbor interactions (all interactions are antiferromagnetic) on a square lattice in two alternative versions: first, $J_1\neq 0$, $J_2\neq 0$, $J_3=0$ and second, $J_1\neq 0$,  $J_2=0$, $J_3\neq 0$. The main goal of this paper was to find and investigate the magnetic properties of these two models, and to compare them to the properties of 1D chain with the nearest-neighbor and second-neighbor interactions $J_1\neq 0$ and $J_2\neq 0$.

Magnetic properties of both versions of the model have much in common. The frustration points in the absence of magnetic field coincide $R=R'=1/2$. The magnetization curves have plateaus and two frustration fields, and the upper fields also coincide. In all the frustration fields the entropy does not vanish at $T\rightarrow 0$. In the case of strong nearest-neighbor interaction and below the lower frustration field the magnetic structure is the same (N\'{e}el antiferromagnetic). At the frustration points and at all the frustration fields the transition temperature is suppressed to zero.

However, substantial dissimilarities in the magnetic properties do not enable the two versions of the model to be referred to as similar. In the absence of magnetic field, at the frustration point $R=1/2$ in the first version the entropy vanishes at $T\rightarrow 0$, while it tends to a non-zero in the second version at identical frustration point $R'=1/2$. The heights of plateaus are different, $1/2$ of the saturation magnetization in the first version and $1/3$ in the second. The lower frustration fields differ. At the frustration points $R=R'=1/2$ the magnetization curve in the first version begins antiferromagnetically, while in the second ferromagnetically. When the nearest-neighbor interaction is weak ($R>1/2$ and $R'>1/2$) the magnetic structures have utterly diverse translational symmetry, SAF in the first version and quadruple in the second. The intermediate structures in between the lower and upper frustration field also have quite different symmetry.

A comparison between the second version of the model (that can equally be called a model with nearest-neighbor interaction and second-neighbor interaction along sides of a square) and the 1D linear chain gives radically different result. Apart from previously established common features many new are revealed. The heights of plateaus are the same, namely, $1/3$ of the saturation magnetization. At the frustration points $R'=1/2$ the magnetization curves begin ferromagnetically. The translational invariance of all the structures (at any value of $R'$ and magnetic field) the second version along either side of a square coincides with that in a linear chain. All the frustration fields coincide being expressed in the rescaled form $h=H/z$. A complete agreement between the magnetizations (Figs.~\ref{fig:5}~and~\ref{fig:11}) is the most striking similarity of the second version of 2D model and a linear chain. 

We may ultimately conclude that magnetic properties of the linear chain with nearest-neighbor and second-neighbor interactions and the 2D model with nearest-neighbor and third-neighbor interactions are alike at almost every aspect provided that the temperature and magnetic field are rescaled. 

We predict that at appropriate choice of a model the magnetic properties in 1D, 2D, and 3D lattices will be similar. In particular, we do believe that the future numerical calculations of magnetizations on the Ising simple cubic lattice with nearest-neighbor interaction and second-neighbor interaction along all three cube edges should reproduce the 1D magnetizations from Fig.~\ref{fig:5}.

\section*{Acknowledgments}
This work was partially supported by projects no.12-I-2-2020 of Ural Division RAS, no.12-P-2-1041 of Presidium RAS, by Russian Foundation for Basic Research (projects  no.12-02-96504-r yug a, and no.13-02-00220).

\end{document}